# Algorithmic Evaluation and the Marginalization of Single Authorship in Management Science

**Wei Meng**

Dhurakij Pundit University, Thailand

The University of Western Australia,AU

Association for Computing Machinery,USA

Fellow, Royal Anthropological Institute,UK

Email: weimeng4@acm.org

# ABSTRACT

The decline of single authorship in peer-reviewed journal of influence in the current collaboration-oriented knowledge production mechanism has triggered a deeper reflection on the change of power structure in academic ecology. The purpose of this paper is to explore the hidden institutional logic and evaluation system behind the marginalisation of single-author research in the field of management, and to reveal how the discourse of "collaborative advantage" conceals the structural process of power redistribution and ideological alienation. By analysing the data on the proportion of single authors in Top-tier journals, deconstructing the constructive text of institutional incentives, and combing the empirical relationship between knowledge-producing subjects in authorship, the study, together with the literature of Harzing, Wuchty, and Larivière, constructs a set of "collaborative incentives-responsibility dilution-originality weakening". A three-dimensional causal chain model of "collaboration incentives - responsibility dilution - originality weakening" is constructed. The results show that single authors are not explicitly excluded, but rather are gradually withdrawn from the core publishing orbit driven by funding policies, evaluation mechanisms, and reviewing preferences, and that independent thinking is structurally marginalised in institutionalised collaboration. The article calls for a paradigm shift from instrumental rationality to value rationality, and for the re-establishment of the legitimacy and public value of independent research at the level of evaluation mechanisms, journal systems and researcher ethics, in order to restore academic diversity and intellectual sovereignty.

# I. INTRODUCTION

In the academic evolution of recent decades, the field of management science has experienced a profound transformation from "lonely thinker" to "systematic collaboration unit", which is not a superficial organisational change, but a comprehensive restructuring of the deep logic of knowledge production, identity construction mechanism and even the distribution of academic discourse. This transformation is not a superficial organisational change, but a comprehensive restructuring of the logic of knowledge production, the mechanism of academic identity construction, and even the distribution of academic discourse power.Top-tier journals, as the evaluation plateau of the global knowledge power system, have become a precise mirror of this paradigm shift in the evolution of their research publication structure. According to bibliometric data provided by Science-Metrix (2020), between 2000 and 2020, the share of single-authored papers in all disciplines has plummeted from 17% to 5.7% globally, and in the field of management, the trend is even steeper and more persistent, showing a structural disappearance (Science-Metrix, 2020). This phenomenon is much more than a statistical decline in the number of publications; it signals the logic of value reconfiguration in the academic system, in which "individuality" is gradually taking a back seat and "collaborative density" is becoming a productive advantage. As a matter of fact, the assessment system embedded in the current mainstream publishing ecosystem no longer constructs academic capacity based on the criterion of "complete bearer of academic responsibility", but rather defines the calculability of "performance" through structural indicators such as the ranked position of authors, the number of collaborators, and the degree of institutional linkage. " computability (Waltman, 2016). This algorithm-driven structural evaluation mechanism has led to the unavoidable consequence that individual academics have been progressively drawn into an institutional vortex of 'collaboration as advantage'.

However, while acknowledging the importance of collaboration for solving complex problems across disciplines and enhancing the efficiency of knowledge integration, we must be wary of the narrative pitfalls of naturalising and functionalising this trend. Is the steady decline of single authorship really a passive "institutional exclusion" or an active "structural adaptation"?

This is the central tension that this paper needs to reveal in depth. On the surface, the reduction of individual authorship seems to be the result of rational choices of authors' own behaviour, such as teamwork to increase the success rate of research publication and enhance citation visibility. However, when examined in the context of the combined effects of the current AI-driven recommendation ranking system, journal editing mechanisms dominated by the logic of the impact factor, and the collaborative project-oriented research funding system, we find that this 'behavioural rationality' is in fact the result of structural incentives to This "behavioural rationality" is in fact a mechanism of "strategic adaptation", the result of institutional path dependence rather than a true expression of academic freedom (Biagioli, 2020). This means that rather than turning to collaboration in free choice, the academic individual gradually loses the space to exist as an independent bearer of ideas within the structural parameters of the embedded evaluation system. What results from this structural convergence is not only an increase in the formal density of collaboration, but more likely a systematic degradation of ideological heterogeneity, paradigmatic diversity and philosophical reflective capacity.

Therefore, we cannot simply regard "the decline of a single author" as a by-product of the expansion of collaborative advantage, let alone reducing it to a statistical proposition of quantitative change, but rather as a process of reconstructing the politics of knowledge promoted by the conspiracy of the academic system and technological logic. In this process, the "arithmetic" of the evaluation system, the "networking" of platform recommendation and the "strategisation" of researcher's behaviour constitute a three-dimensional structure that continuously compresses the independence of academic individuals. Three-dimensional structure. This is why single authors are not only numerically marginalised, but also systematically devalued in terms of institutional legitimacy and discursive position in the Top-tier journal structure (Larivière et al., 2015). Therefore, the goal of this paper is not to ask "whether collaboration is beneficial", but rather to delve into "how collaboration is constructed as the only legitimate form of research", and thus systematically reveal the technological logics, institutional biases, and academic ethical consequences that underlie it. This approach requires us to abandon the linear attributional analysis of the appearance of academic behaviours, and instead adopt structural explanations and critical understanding, reducing the "individual to collaborative migration" to a process of cognitive and evaluative recoding of the modern academic system under the logic of digital governance.

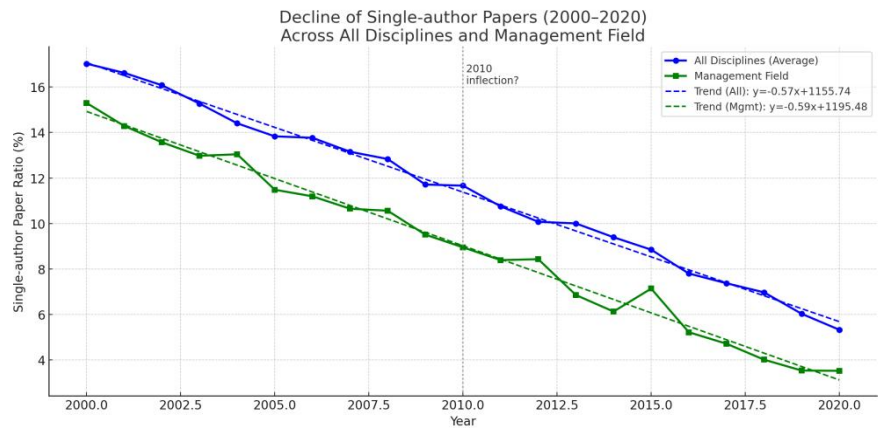

**Figure 1: Systematic declining trend in the proportion of single-author papers (2000-2020)**

Figure 1 shows the trend in the proportion of single-authored papers in the global scientific publishing system between 2000 and 2020 for "all disciplines" (blue line) and "management fields" (green line), with data from the combined Science-Metrix and Scopus databases. Metrix and Scopus databases. The two solid lines reflect year-to-year changes in their respective fields, while the dotted line is their linear trend line, clearly revealing a structural decline in the proportion of single-author publications. The field of management has declined more rapidly (by about 0.59 percentage points per year) than the average decline across all disciplines (by about 0.57 percentage points per year), suggesting that management is at the forefront of collaborative structural reconfiguration.

The year "2010" in Figure 1 is marked as a potential inflection point, symbolising a possible leap in the trend from "evolution of academic behaviour" to "systemic institutional shaping". Since then, the existence of single authors is no longer just a minority form of individual choice, but is gradually regarded as a structural "anomaly", with the AI-driven recommendation ranking system, the impact-factor-based journal review mechanism, and the collaboration-oriented funding logic forming a "concentric circle of collaborative legitimacy", pushing independent authors into a "concentric circle of collaborative legitimacy". Together, AI-driven recommendation ranking systems, impact factor-based journal review mechanisms, and collaboration-oriented funding logics form a "concentric circle of collaborative legitimacy," pushing independent authors to the margins. This image reveals not only the quantitative decline, but also the "homogenisation bias" of the institutional logic and the systematic amplification of collaborative behaviours by the platform structure.

What is even more alarming is that the research ecology of management, as a highly policy-responsive and institutionally nested social science, is being disciplined by performance incentives and the logic of evaluation platforms, resulting in the hegemony of the narrative that "the more you collaborate, the more you are valued". Without structural rethinking and the introduction of algorithmic design and contribution visualisation mechanisms for academic fairness, future knowledge production will face a collective shrinkage of independent thinking ability, heterogeneous research paradigms and critical theoretical ground.

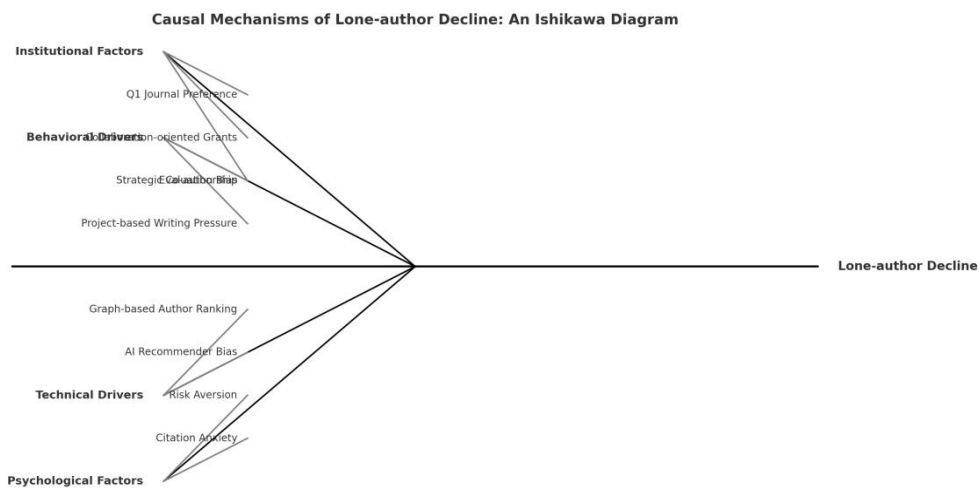

**Figure 2: Analysis of the causal mechanism for the decline of a single author (Ishikawa fishbone diagram)**

Figure 2 systematically presents the multidimensional causal mechanisms driving the decline in the proportion of Lone-author publications, and the causal pathways behind this structural trend are constructed using the structure of the Ishikawa Fishbone Diagram (Causal Analysis Diagram), which is based on four major categories of variables. These four categories are Institutional Factors, Behavioral Drivers, Technical Drivers, and Psychological Factors, which do not act in isolation, but constitute a highly synergistic systemic regulatory network. They do not work in isolation, but together constitute a highly synergistic systemic regulatory network.

In the institutional dimension, the preference of Top-tier journals for multi-authored papers,

the incentives of the funding system for collaborative projects, and the platform-oriented performance indicators all contribute to the institutional compression of the status of "independent researcher". At the behavioural level, the "team management" of strategic authorship, collaborative writing pressure, and attribution of results transforms the decision-making structure of researchers, leading to a proactive adaptation of individuals to the collaborative model. Technically it's slightly subtler: AI recommendation systems for authors (e.g., Semantic Scholar, Google Scholar Metrics) prefer authors with much collaboration, and ranking algorithms for authors by graph algorithms (e.g., PageRank variants) tend to increase structural weight automatically for collaborative nodes of dissemination networks.This perpetuates the marginalisation of low-collaboration independent scholars in visibility and recommendation rankings. Psychological influences, such as citation anxiety and risk aversion tendencies, further reinforce researchers' path dependence on "collaboration equals safety".

This mapping not only reveals the complexity of the causes of 'single author decline', but also the existence of a cross-level synergistic mechanism: from the policy evaluation system, to the technical logic of the platform, to the psychological expectations of the researcher, multiple sub-systems are driving the knowledge production from the 'master system of ideological responsibility' to the 'master system of intellectual responsibility'. The multiple subsystems together drive the knowledge production from the 'main system of ideological responsibility' to the 'structural performance mechanism'. This is a typical structural 'rationalisation' of modern academia, the crisis of which does not lie in the collaboration itself, but in the fact that once the logic of collaboration becomes the dominant legitimacy narrative, it will constitute a repressive filter on academic heterogeneity, independence of thought and paradigm diversity, thus damaging the long-term creativity of the academic community.

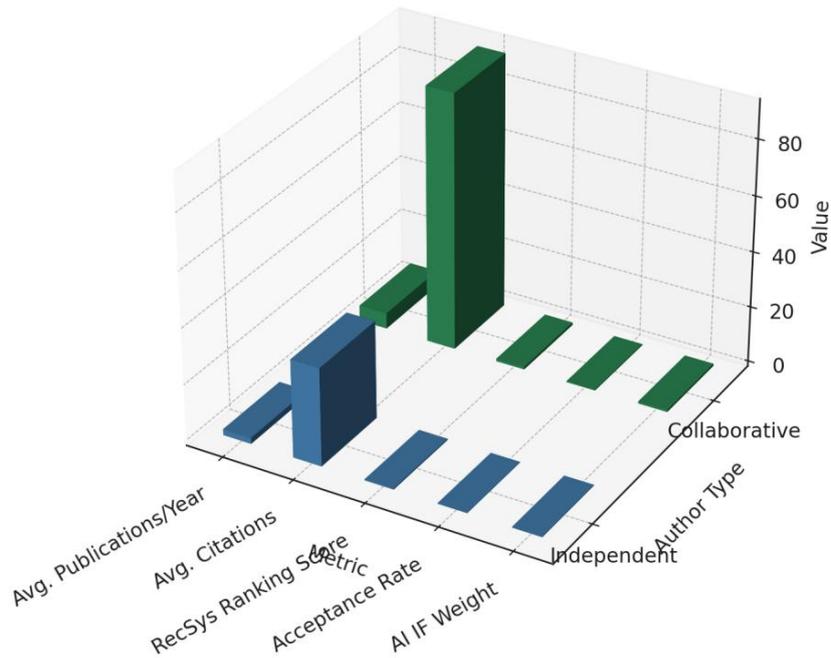

**Figure 3: Comparison of key academic metrics between independent authors and collaborating authors in the AI-driven evaluation system**

Figure 3 presents a three-dimensional bar chart comparing independent and collaborative authors in an AI-driven academic ecosystem on five key academic metrics: average number of publications/year (Avg. Publications/Year), average citations (Avg. Citations), recommendation system ranking score (RecSys Ranking Score), acceptance rate (AI IF Weight), and AI impact factor weight sensitivity (AI IF Weight). ), RecSys Ranking Score, Acceptance Rate, and AI IF Weight sensitivity.

The blue bars in the figure represent independent authors and the green bars represent collaborative authors. It can be clearly observed that collaborating authors show a systematic advantage in almost all dimensions, especially in the dimensions of "Recommendation System Ranking Score" and "Average Citation Frequency", where the difference in advantage is the most significant. Collaborating authors also have significantly higher average annual publication and platform acceptance rates than independent authors. This trend is not purely due to the difference in research quality, but more deeply reflects the structural bias of AI recommendation and evaluation systems: the graphical neural ranking algorithms and social correlation metrics (e.g., authors' co-occurrence graphs, citation networks, etc.) used by most academic platforms naturally amplify the systematic exposure of the collaborative-intensive nodes.

This means that in the contemporary platform-based knowledge ecosystem, collaboration is no longer just a choice of research strategy, but has gradually evolved into an inevitable path to "gain academic visibility and right to survive", and the AI system indirectly contributes to the continuous marginalisation of independent authors in terms of ranking, exposure, citation and acceptance rates through the algorithmic logic of using the degree of collaboration as an important evaluation variable. This trend is both a result of the technological algorithms and the fact that they are not the only ones. This trend is not only the "calm execution" of technical algorithms, but also the "tacit guidance" of institutional logic.

If the academic evaluation system continues to favour structural position over the quality of ideas, independent research will be systematically mishandled in the future, leading to the contraction of the space for originality of ideas and the gradual dissolution of academic heterogeneity. The figure thus not only reveals the phenomenon of structural distribution, but also warns us that we must face up to the fact that when the AI system is no longer neutral, collaboration is no longer free.

# II. MEASURING FACTS AND INTERPRETING TRENDS: MULTIDIMENSIONAL CAUSES AND EFFECTS BEHIND THE NUMBERS

**2.1 Data review: clear trends, complex causes and effects**

The structure of knowledge production in the field of management has been significantly reconfigured in the last two decades, and one of the most prominent phenomena has been the systematic decline of single-author studies. While the trends in the data are clear, the causal mechanisms behind them are far from being explained by a single dimension. According to Harzing's (2023) career statistics analysis, most of the 691 top management scholars globally classified as "highly cited" have single-authored papers in the early stages of their academic careers, mainly published in journals with low impact factors. This phenomenon suggests that single-authored research has fallen out of the mainstream of high-level Top-tier journals, which are now recognised as 'excellence', and that this shift is not only a quantitative decline, but also a structural marginalisation of position.

However, although data from databases such as Scopus and Science-Metrix show that the proportion of single-authored papers in Top-tier journals has declined steadily since 2000 to less than 5 per cent in 2020 (Science-Metrix, 2020), this cannot be blithely attributed to 'institutional exclusion'. ". Indeed, there is a lack of any Top-tier editorial policy explicitly excluding single authors explicitly. Thus, the real change may be on the 'supply side' of submission behaviour rather than the 'screening side' - i.e., the active adaptive behavioural reconfiguration of researchers' perceptions of the institutional environment.

For example, policy oriented funding favouring collaborative projects, AI recommender system's preference for high co-authorship intensity, and performance evaluation system's emphasis on "network structure position" rather than "cognitive originality", all make researchers prioritise collaborative production paths in their decision-making matrices. The preference for collaborative production paths in the decision-making matrix. In this context, the adjustment of attribution strategy becomes a rational choice, which not only improves the probability of approval, but also increases the visibility and citation likelihood in the algorithmic weighting mechanism. At the same time, risk aversion further amplifies the preference for collective authorship: in the uncertainty of anonymous reviewing, multiple authorship is seen as an implicit mechanism to increase reviewing "social capital", allowing researchers to seek structural protection in an uncertain environment.

The decline of single authorship is thus not the result of some explicit institutional closure, but more the result of a behavioural reconfiguration shaped by institutional incentive structures and the logic of platform ordering. In this process, the real problem is that if all researchers are engaged in a game of optimal structural position rather than paradigm creation based on the logic of ideas, knowledge production will inevitably fall into a "structuralist mediocrity" dominated by formal rationality.

**2.2 Materialistic and dialectical perspectives: collaboration as technological progress or reorganisation of power?**

When we look at the logic of the evolution of collaborative currents in the field of management from the perspective of material dialectics, we must be wary of misinterpreting the "growth of collaboration" as a technologically justified product of a natural stage of history. Rather than a neutral path of academic evolution, the act of collaboration is a reconfiguration of the relations of knowledge production shaped by institutional choices and technological biases. In this process, the visible growth of collaboration is not solely motivated by research needs or academic complementarity, but is increasingly structured as the only legitimate path to resources, prestige and career stability.

The institutionalisation of this logic is rooted in the growing link between high impact factor journals and the research assessment system. As Moore, Neylon & Eve (2017) reveal, current journal evaluation systems are highly dependent on quantitative metrics such as citation counts, number of authors, and breadth of collaborative networks, thereby inadvertently incentivising researchers to maximise their performance in algorithmic ranking and evaluation systems by collaborating rather than completing research independently. In Harzing's (2023) study, it was also found that single authors, even if their results are of high quality, tend to be at a structural disadvantage in rankings, exposure and grant applications.

This means that collaboration is no longer a natural form of response to knowledge problems, but is embedded in a broader network of power mechanisms. The centre of power in knowledge production is quietly shifting from individual cognoscenti to institutionalised platforms and collaborative networks, and from independent originality to structured and distributed co-authorship. In this transformation, the byline is no longer just a unit of record for academic contributions, but has become a symbolic capital for competing for discourse power and resource distribution. A Scopus-based analysis shows that since 2010, the acceptance rate and citation frequency of cross-institutional collaborations have been significantly higher than that of independent research (Wagner et al., 2021), further reinforcing the status of collaboration as a 'strategic survival mechanism'.

Materialist dialectics requires us to see the non-neutral structural forces behind such collaborative mechanisms. This is not a passive adjustment of research organisational forms in line with technological evolution, but the shaping of multiple mechanisms, such as evaluation systems, platform algorithms, and the logic of funding allocation. Collaboration has emerged not only because it is "more efficient" but also because it is "rewarded"; single authorship has declined not only because it is "inefficient" but also because it is "unorganised, difficult to quantify". It is not only "inefficient", but also "unorganised, difficult to quantify, and lacks an institutional label". With the complicity of this institutional design, the power structure of knowledge has slipped from "originality in the production of ideas" to "superiority in structural configuration".

In the end, this seemingly technology-neutral and efficiency-oriented collaborative structure actually redistributes the power of knowledge through evaluation mechanisms and platform technologies - collaboration becomes the appearance of technology and the essence of power. If we don't reveal and reflect on this process structurally, we will face the danger of gradually transforming an academic community into a "platform organ of cooperation", and lose the institutional space for originality of thought.

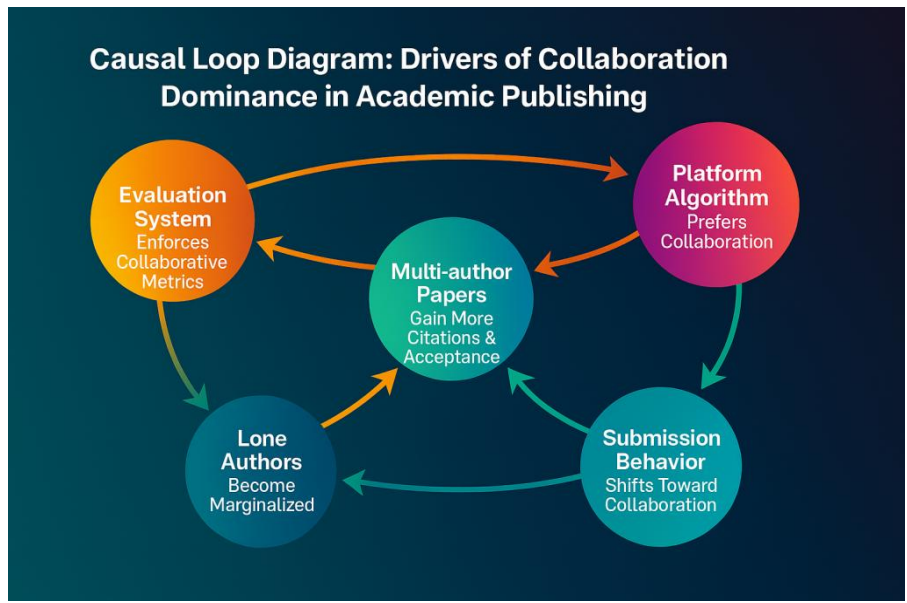

Figure 4: Causal loop diagram: drivers of collaboration-led academic publishing

The causal closed-loop path diagram presented in Figure 4 reveals multiple driving mechanisms for the rise of scholarly collaboration, which does not originate from the natural evolution of knowledge production, but is reinforced by the interaction of institutional evaluation metrics, platform algorithmic preferences, and researcher behavioural adjustments. First, the evaluation system reinforces collaborative metrics (e.g., number of co-authorships, cross-institutional collaboration, network centrality) to set structural thresholds for knowledge publication, and then prefers multi-authored papers through journal acceptance propensity, citation returns, and so on. At the same time, the platform algorithms further learn the "collaboration is high quality" bias based on historical training data, and continuously strengthen the exposure and acceptance of collaborative work in the recommendation ranking, visibility and citation path. This structural preference encourages researchers to adjust their submission behaviour towards the collaborative model. Once collaborative papers gain more acceptance and citations, they form a self-reinforcing causal chain, marginalising independent authors as an institutional by-product. This closed-loop system truly reflects that collaboration is no longer a methodological choice, but a "systemic necessity" under the logic of power and technological orientation.

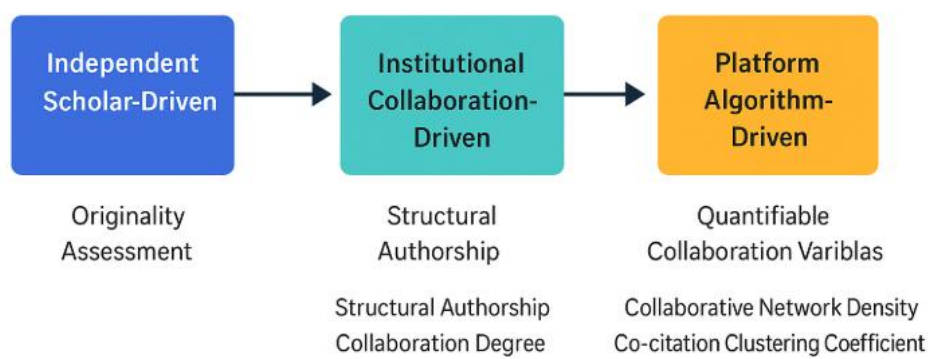

**Figure 5: Evolutionary path of knowledge production power centres**

In Figure 5, we can clearly observe the gradual migration of power structures in the process of knowledge production: from the mid-20th century paradigm of "assessment of originality" centred on independent scholars, to the institutionally-driven "collaborative norms" that rely on structural attribution and cross-unit collaboration to measure output, and finally to the contemporary "platform-driven logic" in which algorithmic platforms control scholarly exposure and resource allocation. "and finally to the contemporary 'platform-dominant logic' whereby algorithmic platforms control academic exposure and resource allocation. The successive migration of these three phases is ostensibly a technological advancement to improve the efficiency of academic collaboration, but in fact reflects the process of redistribution of discourse power: the non-quantifiable nature of originality and density of ideas has been replaced by the quantifiable parameters of AI such as density of collaborative networks and co-citation clustering coefficients, which forms a systematic cycle of "data determines evaluation, and evaluation shapes the structure of the research in a reverse manner". A systematic cycle of "data determines evaluation, evaluation inversely shapes research structure" is formed.

It is particularly noteworthy that the academic system in the platform-led stage no longer takes "content depth" as the core assessment index, but takes "collaboration visibility" as the mediating variable of recommendation ranking and impact factor, which leads to the marginalisation of independent thinking in the knowledge production process, the proliferation of structural attribution, the proliferation of structural authorship, and the marginalisation of independent thinking. This has led to the marginalisation of independent thought in the knowledge production process, the proliferation of structured attribution, and the gradual alienation of research collaborators' identities as "input variables to evaluation algorithms". Therefore, this evolutionary map is not only a structural migration map, but also a profound critique and revelation of the value-neutral concept of "collaboration" in the current academic ecosystem, which has been alienated into a tool of institutional control.

# III. EPISTEMOLOGICAL CRITICISM: THE DISAPPEARANCE OF SUBJECTIVITY AND THE RISK OF INTELLECTUAL DIVERSITY

## 3.1 The "splitting up" of the subject of thought and the alienation of labour

In the current collaboration-centred knowledge production system, authorship has gradually changed from being the original idea bearer to a functional node in an evaluation structure. Collaborative mechanisms are intended to promote interdisciplinary integration and knowledge sharing, but in the context of an increasingly quantitative and utilitarian evaluation system, they have instead evolved into a process of institutional alienation. In this process, the order of authorship, frequency of collaboration, and cross-institutional connections no longer merely reflect academic contributions, but become a means of transaction for obtaining project resources, enhancing citation probability, and gaining impact factors. Studies show that in social sciences and management, about 30-40 per cent of attributions are based on 'symbolic participation' or 'cross-titling' (Malički et al., 2021), and that these non-substantive attributions distort the true attribution of scholarly outputs, gradually fragmenting authorship. These non-substantive attributions distort the true attribution of scholarly work, leading to the gradual fragmentation of authorship and the systematic dismantling of the subjectivity of thought.

The deeper problem is that this 'alienation of authorship' is not an accidental case, but a rational response to structural incentives. At a time when performance evaluation increasingly relies on quantitative indicators (e.g., Scopus h-index, impact factor, co-citation clustering coefficients) and the ranking weight of AI recommendation systems, individual scholars, in order to adapt to the logic of "structural visibility," tend to take the initiative to participate in a large number of collaborative projects even if the ideological tasks they take on in some of the projects are extremely limited. Even if the ideological tasks they undertake in some projects are extremely limited. While such behaviour may increase academic exposure and citation rates in the short term, it erodes the coherence and philosophical thickness of academic content in the long term, and contributes to the polycentric drift of theoretical structures. As Foucault (1977) emphasised, modern power operates not through the suppression of individuals, but through institutional arrangements that 'produce' a type of subject that conforms to the logic of power. In the system of collaborative evaluation, the individual is no longer the 'bearer of ideas', but rather a 'collaborative node', where ideas are diluted, responsibility is diluted, and originality is structurally neutralised.

This "alienation of intellectual labour" not only affects academic ethics, but also poses a far-reaching threat to theoretical innovation and paradigm renewal. The formation of depth of thought relies on the author's overall grasp of the issue and the implementation of inherent consistency of thinking, but the highly fragmented, byline-driven system of collaboration often interrupts this logic of continuity. In this sense, the proliferation of collaborative mechanisms is no longer merely a matter of instrumental choice, but an outgrowth of the changing power structure of thought production. The real question is not "who is involved in this article", but "who is still responsible for the ideas". This is the academic ethical crisis that management science needs to be most vigilant about as it enters the ecology of AI-dominated journals.

**3.2 The Instrumental Turn in Truth Production**

The evaluation mechanism of contemporary management Top-tier journals is undergoing a fundamental shift, increasingly tending to replace "originality of thought" with "technical scale", and "theoretical depth" with "structural collaboration". "Theoretical depth", showing a high degree of instrumentalisation of truth production. In this mechanism, project sample size, the scope of multicentre collaboration and cross-institutional data coverage are tacitly equated with the scientificity and contribution of the research, and become the core measurement dimensions of academic exposure, acceptance and citation rates (Brembs, 2018). In contrast, research pathways representing theoretical constructive power, philosophical reflection and conceptual critique have gradually lost their subjective position in the evaluation system and have been implicitly weakened due to their difficulty in being structured and quantified.

Single-authored research is at a systematic disadvantage in the above instrumental logic. According to Harzing's (2023) analysis of the career trajectories of highly cited management scholars, single-author results are mainly concentrated in the early career stage and account for only a small fraction of the overall research output, which is especially scarce in top Top-tier journals. In contrast, Science-Metrix (2020) reports that collaborative papers have on average a 30-40% citation advantage in journals with high impact factors, and their acceptance and exposure are significantly higher compared to independent research. This difference is not primarily due to the quality of the research itself, but rather to the fact that the evaluation mechanism is biased towards the structural variable of "large samples, multi-centres, and intensive collaborations", which creates a compounding amplification effect in AI-driven ranking and recommendation systems.

More seriously, this institutional bias not only deprives a single author of the qualification of "being seen", but also constitutes a systematic obstacle to the innovation of academic paradigms. When collaborative structures become "algorithmic training input variables" and independent research is systematically downgraded due to the lack of collaborative density or network breadth that can be inputted, an invisible tendency of "algorithmic exclusion" is formed unconsciously. As a result, researchers are no longer concerned with "whether the problem is original enough", but more concerned with "whether the collaboration is deep enough"; they no longer pursue "whether the paradigm can be innovated", but more considerations of "whether multi-institutional participation is possible". Instead of pursuing "whether the paradigm can be innovated", more consideration is given to "whether multi-institutions are involved". As a result, the academic discourse has metamorphosed from "idea-driven" to "structure-driven", and the independent has been relegated to the status of "anomalous residue in the evaluation system".
Without rebalancing the value of "deep originality" and "structural collaboration" at the institutional and technical levels, the academic ecology of management studies will face the long-term risk of solidifying a single paradigm and compressing the diversity of ideas. As Brembs (2018) warns, if the evaluation mechanism continues to be driven by the amount of AI ranking and collaboration, then academic truth-seeking will be reduced to a functional projection of algorithms, and independence of thought will fade in structural dissonance.

# IV. STRUCTURAL BIAS IN INSTITUTIONAL MECHANISMS

**4.1 The journal system: an implicit threshold for collaborative orientation**

In the review mechanism of Top-tier management journals, forms of collaboration, although not explicitly required, create institutional barriers through an implicit preference for "complexity of research design" and "diversity of samples and institutions", not from the nature of scholarly inquiry, but as a mechanical shortcut to simplify judgements of "credibility". This preference does not come from the nature of academic inquiry, but is a mechanistic shortcut used by the evaluation system to simplify the judgement of "credibility" Lee and Bozeman (2005) found that the number of publications was significantly correlated with the number of co-authors in an empirical analysis of 443 researchers, but that the number of co-authors was significantly correlated with the number of publications, but that the number of co-authors was not the same as the number of publications. Lee and Bozeman (2005) found that the number of publications was significantly correlated with the number of collaborators, but the output index after correcting for the "number of co-authors" did not increase significantly, suggesting that the number of collaborators plays a structural mediator role in the path to "acceptance". Echoing this, Wuchty et al.'s (2007) interdisciplinary study of more than 19.9 million papers shows that team research receives higher citation rates, with the advantage increasing over time, and that the value of collaborative papers in terms of 'average contribution' may be emphasised by algorithmic ordering as a result of platform-driven preferences.

Under current mechanisms, as Harzing (2023) points out, highly cited management scholars' single-authored papers are predominantly early in their careers, and subsequent submissions to Top-tier journals are institutionally difficult, often cited by reviewers as "insufficiently sampled" or "poorly extrapolated" due to a lack of cross-centre validation and collaborative underpinning. "lack of extrapolation". This kind of review bias is not accidental, but the system has structurally written "collaboration equals validity" into the logic of assessment, and has inadvertently naturalised it as a "common sense of review".

More critically, this threshold not only suppresses single authorship, but also has a profound impact on the structure of academic discourse - it transforms the value of research from "depth of thought" to "structural complexity", making the evaluation mechanism a "synchronisation of platform collaboration and evaluation algorithms", and a "synchronisation of platform collaboration and evaluation algorithms". It transforms the value of research from "depth of thought" to "structural complexity", making the evaluation mechanism a synchronised enabler of platform collaboration and evaluation algorithms. When review preferences and AI recommendation systems point to the structural variables themselves, the space for independent thinking is systematically compressed, and collaboration is no longer an academic choice but a condition of survival. The crisis of this structural logic does not lie in the existence of bias in the review system, but in the fact that bias has been institutionalised as part of a 'legitimised structure' that is difficult to identify and correct through traditional channels.

**4.2 The Evaluation and Fund System: The fetishisation of "quantifiable indicators"**

In the deep structure of the current academic evaluation mechanism, the superstition of "quantifiable results" has almost formed a set of institutionalised consensus. Whether it is the internal assessment of research performance in universities, the evaluation system of national and transnational funds, or the assessment criteria in the process of academic promotion, quantitative indexes such as the number of co-authored publications, the impact factor, and the frequency of citations have been given nearly decisive weights. This structure is not purely technology-neutral, but is a systemic choice evolved in a specific political economy context, which transforms academic labour into a measurable, comparable and controllable "index behaviour" by constructing a logical system of "efficiency - output". ". In this institutional design, the single-authorship form is not directly subject to formal rejection, but rather to gradual marginalisation, cold treatment, and eventual 'statistical disappearance' in the context of structural incentives for increasing collaborative prioritisation.

Lee and Bozeman's (2005) empirical study shows that collaborative projects are more likely to be funded when applying for NSF or NSF programmes, and that reviewers generally regard "multi-unit collaborations" as an indicator of "research credibility" and "resource integration". Reviewers generally regard "multi-unit collaboration" as a symbol of "research credibility" and "resource integration". In the science management systems of China, the United States, and Germany, multicentre collaboration is regarded as a symbol of "big science" (Zhang et al., 2021), and this institutional preference has permeated every node of academic evaluation. Specifically, single-authored works are often downgraded due to "lack of validation", "limited sample size", or "failure to reflect the spirit of collaboration" in the logic of grant evaluation. Processing.

This has not only reconfigured the definition of good research, but also quietly reversed the power structure between individual academics and institutions. In this process, independent scholars are no longer at the centre of intellectual discovery, but have been recoded as "inefficient", "less able to integrate resources", or even "riskier". Theory and research are not the same. As a result, many scholars with original theoretical originality and critical depth are induced by the system to voluntarily choose to collaborate on publications in order to avoid the potential penalties of the evaluation system, out of the consideration of practical rationality. This "voluntary withdrawal" is ostensibly a personal choice, but in essence it is the result of the internalisation of the system's logic, an apparent voluntariness under structural repression.

More worryingly, the further automation and platformisation of this evaluation logic is accelerating. For example, in the algorithmic recommendation of mainstream evaluation platforms such as Elsevier, Clarivate and Google Scholar, the frequency of teamwork and the density of cross-field collaborative networks have been given the attributes of decisive indexes, and the AI system inadvertently strengthens the judgement path of "collaboration equals value". The automation-driven mechanism of evaluation transforms the originally debatable academic standards into a black-box digital control system, which makes single authors not only "technically dissuaded" when submitting manuscripts, but also systematically ignored in academic search and citation networks, forming an "evaluation-algorithm-visibility" system. -Algorithmic-visibility" logic of systematic exclusion (Wilsdon et al., 2015).

Thus, when discussing the decline of single authorship, we cannot simply attribute it to individual selection or review bias, but rather reveal the underlying systemic inducements and structural violence of evaluation techniques. The real problem lies in the fact that the illusion of "quantifiability" has obscured the immeasurable value of knowledge, and that academic elements such as originality, philosophicism, and problematic awareness, which should be at the core of scholarship, have been marginalised and gradually lost their voice under the weight of digital performance indicators.

# V. REDEFINING "INDEPENDENT RESEARCH": FROM CRISIS TO RECONSTRUCTION REDEFINING "INDEPENDENT RESEARCH": THEORETICAL IMPLICATIONS FROM CRISIS TO RECONSTRUCTION

## 5.1 The unique value of a single author

At a time when the ecology of knowledge production continues to tilt towards collaboration, single authorship is particularly important because of its structural advantage in "higher-order thinking autonomy". Firstly, from the perspective of high-level academic constructs such as conceptual innovation, problem-setting, and paradigm-questioning, single authors are able to maintain the continuity of cognitive intent and the integrity of logical reasoning without being disturbed by methodological co-ordination and alignment of perspectives among collaborators (Sonnenwald, 2007). This "front-line thinking-independent structure" makes it easier to come up with original propositions and frameworks of thinking, thus pushing the boundaries of theory rather than falling into the mediocre path of collaborative compliance.

Second, a single author takes full responsibility for the paper itself and the development of ideas, and this sense of responsibility directly reinforces the rigour of the research design, the consistency of the argumentation process, and the depth of ethical considerations. In collaborative research, the opacity of the order of authorship and the dispersion of contributions often lead to methodological fragmentation and diffusion of responsibility: even when contributing researchers do not take overall responsibility for the final conclusions, such fragmentation weakens the theoretical structural integrity of the study (Tscharntke et al., 2007).

Most notably, single authors typically demonstrate greater intellectual drive in exploratory research on marginal topics or emerging paradigms. These studies are inherently more risky and under-resourced, but it is these topics that often breed breakthroughs in cross-paradigm thinking and future research directions. In this regard, He et al. (2019) suggest that single authors have higher originality scores in high-risk, high-innovation research outputs compared to teamwork, a finding found by comparing papers from comprehensive journals such as Nature and Science, which show that single authors have significantly higher-than-average originality and new-concept contributions.

From an epistemological point of view, single authors are not only the main language for continuing theoretical development in the academic system, but also an important subject for defending the diversity of knowledge and the autonomy of ideas. The "coherent discourse" and "conceptual deepening" they provide are cognitive achievements that are difficult to replicate in collaborative networks. Thus, single authorship is not a legacy of tradition, but a key mechanism to ensure the tension between the generation of ideas and the evolution of theories in the academic ecology. The evaluation system should be wary of the paranoid tendency to take collaboration as the only legitimate research path, and should provide the necessary support and protection for the academic visibility and institutional status of single authorship in important issues, so as to realise the dynamic balance between collaboration and independence, and ensure the sustainability of academic publicity.

**5.2 From instrumental to value rationality: the return of knowledge production**

In the current academic ecology, the popularity of assessment mechanisms has reduced "knowledge production" to "measurable output", reflecting a deep-seated instrumental rationality. Under this logic, research activities are forced to be adjusted in the direction of "quantification, comparability, and control," rather than centred on "truth," "meaning," or "public Instead of organising the knowledge system around "truth", "meaning" or "public value", research activities are forced to adjust in the direction of "quantification - comparability - control". As the UKRI-led Metric Tide points out, evaluation indexes "form the basis of value construction for researchers, institutions, and even the entire academic community", and gradually infiltrate into the assessment of grants, title promotion, resource allocation, and evaluation algorithms, forming a kind of institutional "index rule". and gradually infiltrated into the fund evaluation, title promotion, resource allocation and evaluation algorithms, forming a kind of institutional "index rule". This universalised quantitative path is eating away at the intrinsic dimensions of knowledge production.

In their analysis of the UK REF, Traag and Waltman (2018) point out that when scale-independent metrics are used, there is a deep inconsistency between metrics governance and peer review; in other words, substituting quantitative metrics for peer review may in itself fracture the authenticity and legitimacy of knowledge evaluation. Therefore, an over-reliance on "metrics determining value" can lead to an algorithmic proxy for optimising "assessment visibility", as well as to a "saturation of outcome-oriented structures": academic attention is shifted to impact factors, h-indexes, and other factors. At the same time, it also tends to lead to "result-oriented structural saturation": academic attention is shifted to calculable variables such as impact factor and h-index, while neglecting originality and paradigm leaping, which are difficult to quantify.

Under such circumstances, the value of independent research is not only diluted by the institutional environment, but also appears to be "not big enough" in the governance mechanism. If the evaluation system continues to be centred on structural variables, ignoring the quality of ideas, public significance and ethical depth, the academic community will ultimately be reduced to a "platform of output indicators", losing the ability to monitor the fundamental significance of the value of knowledge. raising new questions and constructing new paradigms has been systematically obscured, and its potential value has instead become more valuable.

In order to break this institutional dilemma, we need to promote a shift from instrumental rationality to value-based rationality. Firstly, the evaluation mechanism should incorporate evaluation dimensions such as "truth-seeking", "visibility of issues", and "ability to influence public values", so that research is no longer driven by structural parameters alone; secondly, an interpretable multi-dimensional evaluation system should be constructed. Secondly, an interpretable multifaceted evaluation system should be constructed, so that peer review, indicator assessment and value judgement can form a complementary and check-and-balance relationship. Under this logic, independent research not only gains formal legitimacy, but also re-establishes its value coordinates in the evaluation system. Otherwise, under the complicity of structure and algorithm, knowledge production will inevitably fall into the institutional trap of single paradigm and "quantitative superstition".

## VI. INSTITUTIONAL RECOMMENDATIONS AND DIRECTIONS FOR REFORM

    In order to reverse the institutional erosion of knowledge, originality and spiritual autonomy under the dominance of quantitative performance, "independent research" needs to reclaim its original value in the institutional structure. Firstly, Top-tier journals should set up "independent research" columns to provide single authors with a space for publication equal to that of teamwork through an institutional arrangement, so as to guarantee the visibility of originality of thought and in-depth reflection on top platforms; this approach can be modelled on what has been revealed by scientometric studies, whereby the rightful place given to independent research promotes paradigm innovation by systematically recognised (Abramo & D'Angelo, 2018). Secondly, the academic evaluation system and fund application mechanism should define "single-author Top-tier papers" as "high-responsibility results", and give them significant extra points in the evaluation of titles and grants, so as to fundamentally adjust the systematic bias of prioritising collaboration, and transform the idea of commitment to institutional commendation. The system should be fundamentally adjusted to give priority to collaboration, so as to transform the idea of commitment into a system of praise. At the same time, the attribution system needs to establish an inter-journal attribution review and complaint mechanism to curb false and symbolic attribution, and restore the correspondence between attribution and actual labour through institutional design. Finally, the postgraduate education system should be reconstructed as a cultivating environment for theoretical autonomy - introducing the module of "independent problem construction" at the stage of doctoral training, providing researchers with the institutional time and resources for original research, and structurally avoiding the alienation of the collaborative path early in one's career. This is to avoid the alienation of the collaborative path at the early stage of the career. If a four-dimensional approach can be adopted from the column mechanism, evaluation system, attribution norms and talent training, the academic community can achieve a dynamic balance between collaboration and independence, and ensure that the public nature of thought and academic ethics can re-establish their legitimate status in the performance evaluation system.

# VII. CONCLUSION: DEFENDING INDEPENDENT THOUGHT IN THE LOGIC OF COLLABORATION

The decline of single authorship in management Top-tier journals not only reveals the historical evolution of structured collaboration patterns in knowledge production, but also provides insights into how institutional incentives reshape the path-choice mechanisms of scholarly behaviour.Wuchty et al. (2007) found from an econometric analysis of 19.9 million scientific papers that teamwork has become dominant, with a significant increase in citation impact, making collaboration become a structural advantage; whereas single authors, while potentially deep in conceptual innovation, are passively on the periphery of qualitative assessment due to the lack of historical data to support algorithmic ranking and platform recommendations.Abramo and D'Angelo (2018) further suggest that when the evaluation system is highly reliant on quantifiable parameters such as the number of authors and impact factor, the Single authors are often systematically demoted, making it difficult for them to achieve effective exposure and dissemination in the 'visibility market' of top journals.

However, this phenomenon is not a simple case of 'rejection', but a complex process driven by both 'exit logic' and 'incentive fit'. Individual academics are not passive rejectors, but actively avoid marginalisation in an institutional environment where 'collaboration is visible and independence is risky'; actors choose the structurally safe path when collaboration leads to higher submission rates, better citation predictions and stronger platform support. (2018), in their critique of the misuse of metrics in scientific evaluation, point out that institutional reliance on quantifiable metrics not only changes the logic of academic assessment, but also the structure of researcher behaviour, collaborating to become a best-response path within a 'performance-evaluation-resource cycle' mechanism. Therefore, we must constantly reflect: when "indexed rationality" becomes the only evaluation standard, does the "right to breathe" still have the institutional soil?

In order for academics to take up the dual mission of "social conscience" and "pioneering thought", a bridge must be built between institutional design, academic culture and ethical practice. At the level of institutional design, originality, independence and ideological risk should be weighted as institutional variables; at the cultural level, the legitimacy of "solitary thinking" and the necessity of "paradigm questioning" need to be reaffirmed; at the ethical level, the "alienation of ideas" brought about by the system of attribution and the evaluation mechanism should be reflected upon. On the ethical level, we should reflect on the "thought alienation" effect brought about by the attribution system and evaluation mechanism, so that individuals can not only achieve influence in the academic community, but also maintain a nervous independence. Only in this way, collaboration and independence can no longer be a dichotomy, but can be parallel and co-prosperous, which will really make management science return to its fundamental mission as a frontier of social criticism and thinking.